\documentclass[aps,prb,twocolumn,superscriptaddress]{revtex4-1}

\usepackage{graphicx}

\usepackage[dvipdfmx]{color}
\begin{document}

\title{Pressure tuning of localization and superconductivity in LaOPbBiS$_{3}$ and La$_{2}$O$_{2}$Bi$_{3}$AgS$_{6}$}



\author{Y.~Yuan}
\affiliation{Department of Engineering Science, The University of Electro-Communications, Chofu, Tokyo 182-8585, Japan}
\affiliation{School of Physics and Electronics, Central South University, Changsha 410083, China}
\author{H.~Arima}
\affiliation{Department of Engineering Science, The University of Electro-Communications, Chofu, Tokyo 182-8585, Japan}
\author{M.~Masaoka}
\affiliation{Fundamental Science and Engineering, The University of Electro-Communications, Chofu, Tokyo 182-8585, Japan}
\author{Y.~Naito}
\affiliation{Department of Engineering Science, The University of Electro-Communications, Chofu, Tokyo 182-8585, Japan}
\author{Y.~Hijikata}
\affiliation{Department of Electrical and Electronic Engineering, Tokyo Metropolitan University, Hachioji, Tokyo 192-0397, Japan}
\author{R. Jha}
\affiliation{Graduate School of Science, Tokyo Metropolitan University, Hachioji, Tokyo 192-0397, Japan}
\affiliation{International Center for Materials Nanoarchitectonics, National Institute for Materials Science, Tsukuba, Ibaraki 305-0047, Japan}
\author{Y.~Mizuguchi}
\affiliation{Department of Electrical and Electronic Engineering, Tokyo Metropolitan University, Hachioji, Tokyo 192-0397, Japan}
\affiliation{Graduate School of Science, Tokyo Metropolitan University, Hachioji, Tokyo 192-0397, Japan}
\author{K.~Matsubayashi}
\email{k.matsubayashi@uec.ac.jp}
\affiliation{Department of Engineering Science, The University of Electro-Communications, Chofu, Tokyo 182-8585, Japan}


\date{\today}

\begin{abstract}
	We report the effect of pressure on the electrical transport properties of the four-layer-type bismuth chalcogenide semiconductors LaOPbBiS$_{3}$ and La$_{2}$O$_{2}$Bi$_{3}$AgS$_{6}$ and present the discovery of a pressure-induced superconductivity. In both compounds, the semiconductorlike behavior concomitant with an anomaly in the more insulating state is gradually suppressed with increasing pressure while the superconductivity develops. The obtained phase diagrams, in sharp contrast with the interplay between the conventional charge density waves and superconductivity, suggest that the enhancement of the superconductivity is due to the disorder near the localization threshold. These results indicate that pressure tuning of the local disorder in four-layer-type bismuth chalcogenides provides an attractive opportunity to study the interplay between disorder and superconductivity.
\end{abstract}


\maketitle

\section{introduction}
	The search for superconductivity in the proximity of collective electronic states has been an intriguing research topic in the field of condensed matter physics. In strongly correlated electron systems, such as heavy fermions, iron pnictides, and high-$T_{\rm c}$ cuprates \cite{Gegenwart,Si,Stewart,Monthoux}, an unconventional superconductivity emerges near a quantum critical point, at which the critical temperature for some phase transitions, such as magnetism or charge density waves (CDWs), is driven toward zero using a control parameter. The playground for the exploration of superconductivity can appear not only in metals but also in low-carrier systems, such as B-doped diamond, Li-intercalated ZrNCl, and thin films\cite{Ekimov,Taguchi,Shahar}. In particular, one of the key ingredients in the superconductor-insulator transition is disorder, which gives rise to an enhancement in the superconducting transition temperature $T_{\rm c}$ near the localization threshold \cite{Kohmoto,Feigelman,Yanase,Burmistrov}. Furthermore, it has been suggested that the formation of localized Cooper pairs, possibly associated with the pseudogap in high-$T_{\rm c}$ superconductors, is realized in the highly disordered regime \cite{Dubi,Feigelman,Yanase,Burmistrov,Gantmakher}.

	BiS$_{2}$-based superconductors have layered crystal structures\cite{Mizuguchi,Mizuguchi_2012,Yazici,Zhai,Kotegawa,Sakai}, which are analogous to those of high-$T_{\rm c}$ cuprates and iron pnictides. On the other hand, the parent compound LaOBiS$_{2}$ is a band insulator without magnetic order; the electron doping that originates through the substitution of O with F leads to a metallic state, and superconductivity then emerges. Owing to the wide chemical variety and the high degree of tunability, a large number of BiS$_{2}$-based compounds have been found thus far. While most BiS$_{2}$-based superconductors can be viewed as conventional $s$-wave superconductors both from theoretical and experimental points of view\cite{Wan,Yildirim,Lamura,Yamashita,Kase,Suzuki}, unconventional features, such as an anisotropic superconducting gap and rotational symmetry breaking, are found\cite{Ota, Hoshi}. Despite substantial research efforts, the origin of superconductivity remains under debate. 
 	
	LaOPbBiS$_{3}$ and La$_{2}$O$_{2}$Bi$_{3}$Ag$_{}$S$_{6}$ crystallize in a tetragonal structure and belong to a broader family of BiS$ _{2}$ compounds\cite{Sun, Hijikata}. According to recent detailed structural investigations, LaOPbBiS$_{3}$ has one rocksalt-type PbS layer sandwiched between two outer LaOBiS$_{2}$-type layers\cite{Mizuguchi_2017}. Additionally, it has been pointed out that Pb/Bi site interchange occurs. Neglecting the presence of a small amount of Pb/Bi site interchange, band structure calculations have revealed that the valence band mainly consists of the S-$p$ orbitals in the PbS layer, whereas the conduction band primarily comprises the Bi-$p$ orbitals in the BiS$_{2}$ layer; thus, this valence band is similar to that in LaOBiS$_{2}$\cite{Mizuguchi_2017,Kurematsu}. It should be noted that hybridization occurs between the S-$p$ bands in the rocksalt PbS layer and the Bi-$p$ bands in the BiS$_{2}$ layer near the Fermi energy in LaOPbBiS$_{3}$, which is in sharp contrast with LaOBiS$_{2}$. In real compounds, the electrical resistivity appears to exhibit a semiconductorlike behavior\cite{Hijikata}. By chemically substituting Se with S in LaOPbBiS$_{3}$, the semiconductorlike behavior is suppressed, and superconductivity with a maximum $T_{\rm c}$ = 1.9~K is induced. Furthermore, $T_{\rm c}$ increases monotonically upon application of pressure and reaches 3.6~K at 0.9~GPa\cite{Jha_2020}. 

	The crystal structure of La$_{2}$O$_{2}$Bi$_{3}$Ag$_{}$S$_{6}$ resembles that of LaOPbBiS$_{3}$ upon replacement of the PbS layer with the same rocksalt-type BiAgS$_{2}$ layers\cite{Hijikata}. In La$ _{2}$O$ _{2} $Bi$ _{3} $AgS$ _{6}$, the transport properties are sample dependent, which is caused by the difference in the carrier concentration. Notably, the metallic sample shows superconductivity at $T_{\rm c}$ $\sim$ 0.5~K and exhibits a hump-like anomaly in the resistivity at $T^{\ast}$ $\sim$180~K\cite{Jha}, which is similar to the case of the possible CDW transition in EuFBiS$_{2}$\cite{Zhai}. With Sn substituting Ag, the anomaly $T^{\ast}$ is abruptly suppressed, and a dome-shaped $T_{\rm c}$ is established in La$_{2}$O$_{2}$Bi$_{3}$Ag$_{1-x}$Sn$_{x}$S$_{6}$\cite{Sn_Jha}. However, the relationship between the anomaly at $T^{\ast}$ and the superconductivity remains unclear due to the lack of a microscopic evidence for the CDW.

	In contrast to chemical substitution, application of hydrostatic pressure to a pristine sample is an ideal strategy to tune the electronic interactions without increasing the level of disorder. In this paper, we report the effect of pressure on polycrystalline LaOPbBiS$_{3}$ and La$_{2}$O$_{2}$Bi$_{3}$AgS$_{6}$ by investigating the resistivity, Hall resistivity, ac magnetic susceptibility, and Hall and Seebeck coefficients. Furthermore, we present phase diagrams accompanied by the enhancement of superconductivity.

\section{experimental details}
 	Polycrystalline LaOPbBiS$_{3}$ and La$_{2}$O$_{2}$Bi$_{3}$AgS$_{6}$ samples were synthesized using a solid-state reaction method that was described elsewhere~\cite{Hijikata,Mizuguchi_2017}. High pressure was generated using a piston-cylinder-type clamped cell, and glycerol was used as the pressure transmitting medium. The pressure at low temperature was determined by the pressure dependence of the superconducting transition temperature $T_{\rm c}$ of lead. The pressure cell was loaded in the cryostat of a Quantum Design physical properties measurement system (PPMS) or a $^{3}$He cryostat. Electrical resistivity and Hall resistivity measurements were performed using a standard four-probe technique. The Seebeck coefficient ($S$) was measured through a steady-state method using chromel/constantan thermocouples. In these transport measurements, thin wires were attached to the sample using gold or silver paste. The ac magnetic susceptibility was measured at a fixed frequency of 211~Hz with a modulation field of $\sim$0.1~mT, and the signal from the pickup coil was detected using a lock-in amplifier.
 
 \section{results and discussion}
 \begin{figure}[t]
\begin{center}
\includegraphics[width=0.5\textwidth]{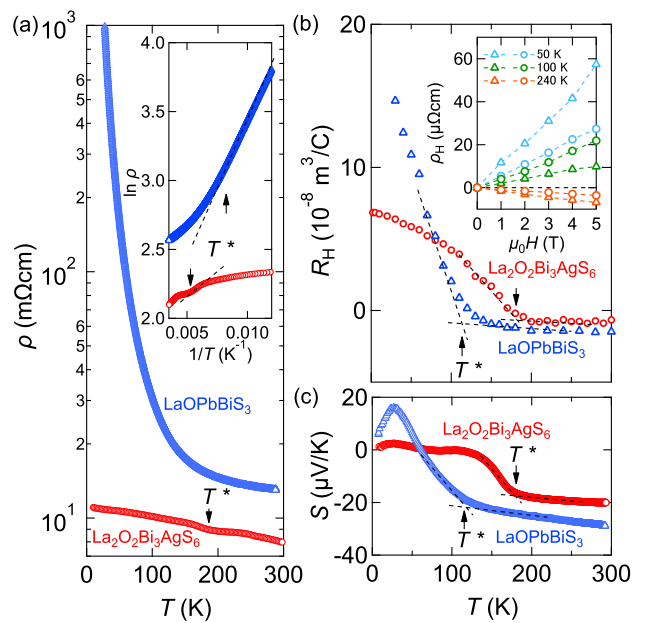}
\end{center}
\caption{(a) Temperature dependence of the resistivity $\rho(T)$ for polycrystalline LaOPbBiS$_{3}$ and La$_{2}$O$_{2}$Bi$_{3}$AgS$_{6}$ at ambient pressure. The inset shows ln$\rho$ vs 1/$T$. Temperature dependence of (b) the Hall coefficient $R_{\rm H}(T)$ and (c) the Seebeck coefficient $S(T)$ for LaOPbBiS$_{3}$ (triangles) and La$_{2}$O$_{2}$Bi$_{3}$AgS$_{6}$ (circles). The inset in (b) shows the Hall resistivity $\rho_{\rm H}$ as a function of the magnetic field at selected temperatures.}
\label{Fig1}
\end{figure}
 
	Figure~1(a) shows the temperature dependence of the electrical resistivity $\rho(T)$ for LaOPbBiS$_{3}$ and La$_{2}$O$_{2}$Bi$_{3}$AgS$_{6}$ at ambient pressure. Both compounds exhibit semiconducting behavior, although the magnitude of $\rho(T)$ for La$_{2}$O$_{2}$Bi$_{3}$AgS$_{6}$ is smaller than that for LaOPbBiS$_{3}$. As mentioned above, the transport properties of La$_{2}$O$_{2}$Bi$_{3}$AgS$_{6}$ are sample dependent due to the extrinsic free carriers arising from the excess Ag, which leads to an increase in the metallic behavior\cite{Jha}. In the La$_{2}$O$_{2}$Bi$_{3}$AgS$_{6}$ metallic sample exhibiting superconductivity, a clear kink is observed at $T^{\ast}$, which is reminiscent of the partial-gap formation associated with CDWs. Our sample, on the other hand, exhibits a nonmetallic behavior in the whole temperature range; $\rho(T)$ increases monotonically with decreasing temperature. Nevertheless, there exists a similar anomaly at $T^{\ast}$~$\sim$~170~K. This indicates that the difference in the transport properties does not have a strong influence on the appearance of $T^{\ast}$. 

\begin{figure}[t]
\begin{center}
\includegraphics[width=0.45\textwidth]{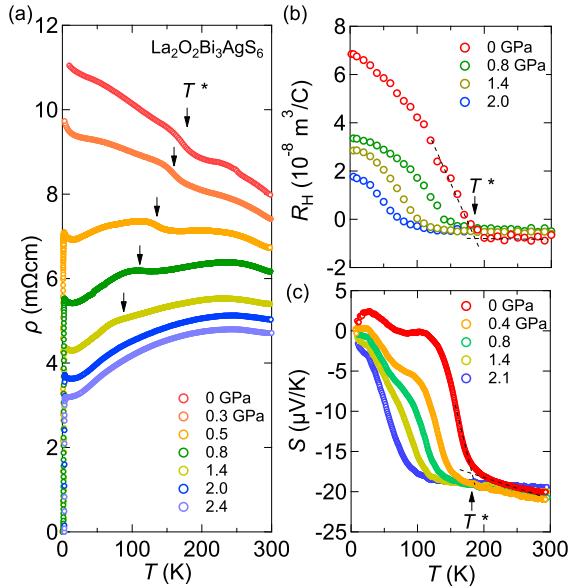}
\end{center}
\caption{Temperature dependence of (a) the resistivity $\rho(T)$, (b) Hall coefficient $R_{\rm H}(T)$, and (c) Seebeck coefficient $S(T)$ of La$_{2}$O$_{2}$Bi$_{3}$AgS$_{6}$ at selected pressures.}
\label{Fig2}
\end{figure}

	The existence of $T^{\ast}$ is also verified by both the Hall ($R_{\rm H}$) and Seebeck ($S$) coefficients, as shown in Figs. 1(b) and 1(c), respectively. The negative values of $R_{\rm H}$ and $S$ above $T^{\ast}$ indicate the dominance of the electronlike conductivity at high temperatures. When the temperature is reduced below $T^{\ast}$, $R_{\rm H}$ suddenly starts to rise and changes sign from negative to positive in both LaOPbBiS$_{3}$ and La$_{2}$O$_{2}$Bi$_{3}$AgS$_{6}$. It should be noted that the magnetic field dependence of $\rho_{\rm H}$ is linear in the temperature range investigated here, as shown in the inset of Fig. 1(b). A similar competition between electron and hole carriers is also observed in $S(T)$, resulting in its remarkable increase below $T^{\ast}$, whereas the complex temperature dependence at low temperatures is ascribed to the phonon-drag effect. In La$_{2}$O$_{2}$Bi$_{3}$AgS$_{6}$, we define $T^{\ast}$ as the minimum in the $d\rho$/$dT$ curve; the value of $T^{\ast}$ thus obtained is almost coincident with that obtained using the criterion illustrated by the dashed lines in the $R_{\rm H}(T)$ and $S(T)$ vs $T$ curves [see Figs.~1(b) and 1(c), respectively]. In the semiconducting LaOPbBiS$_{3}$, the change in the slope of the Arrhenius plot corresponds to the anomaly in $R_{\rm H}$ and $S$ [see the inset of Fig.~1(a)]. As discussed subsequently, a clear deviation from thermal activation behavior is seen at lower temperatures. The coincidence of the $\rho(T)$, $R_{\rm H}(T)$, and $S(T)$ anomalies suggests that $T^{\ast}$ is an intrinsic behavior inherent to LaOPbBiS$_{3}$ and La$_{2}$O$_{2}$Bi$_{3}$AgS$_{6}$. 
	
\begin{figure}[t]
\begin{center}
\includegraphics[width=0.45\textwidth]{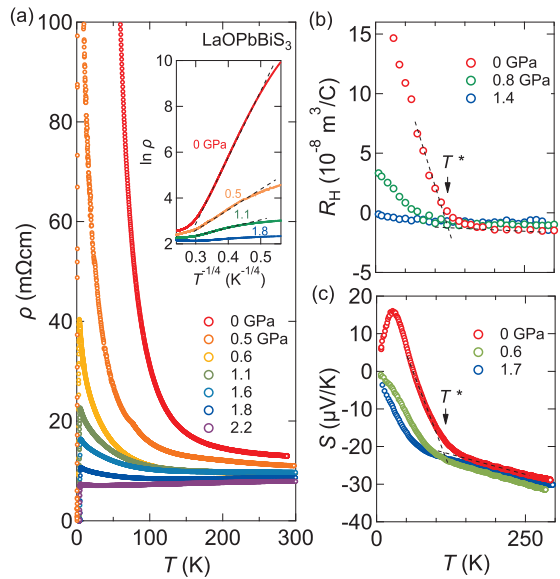}
\end{center}
\caption{Temperature dependence of (a) the resistivity $\rho(T)$, (b) Hall coefficient $R_{\rm H}(T)$, and (c) Seebeck coefficient $S(T)$ of LaOPbBiS$_{3}$ at selected pressures. The inset in (a) shows ln$\rho$ vs $T^{-1/4}$ at selected pressures. The lines are visual guides.}
\label{Fig3}
\end{figure}

\begin{figure}[t]
\begin{center}
\includegraphics[width=0.48\textwidth]{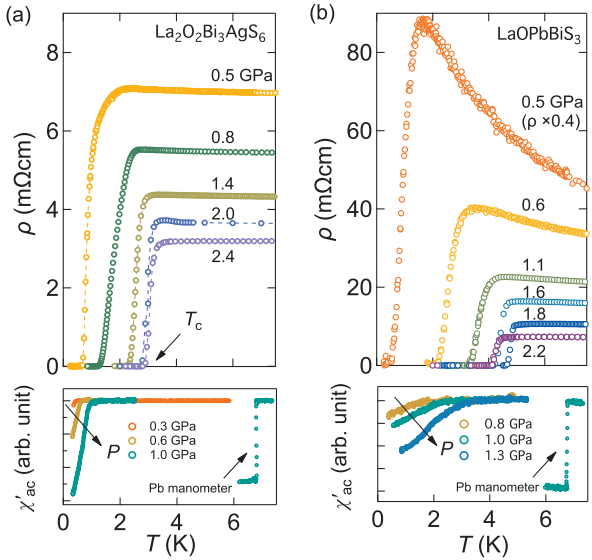}
\end{center}
\caption{Resistivity (upper panel) and ac magnetic susceptibility (lower panel) for (a) La$_{2}$O$_{2}$Bi$_{3}$AgS$_{6}$ and (b) LaOPbBiS$_{3}$ at selected pressures. The superconducting temperature $T_{\rm c}$ is defined as the temperature of zero resistance and the onset of the shielding effect. The signal from a Pb manometer with almost the same volume and shape as the sample is also shown to estimate the magnitude of the superconducting shielding fraction.}
\label{Fig4}
\end{figure}

	Next, we investigate the effect of pressure on the transport properties of these compounds. Figure~2(a) shows the temperature dependence of the electrical resistivity $\rho(T)$ of La$_{2}$O$_{2}$Bi$_{3}$AgS$_{6}$ for pressures ranging from ambient pressure to 2.4~GPa. With increasing pressure, the semiconductor-like temperature dependence is gradually lost, and the anomaly at $T^{\ast}$ shifts toward lower temperatures. Upon increasing the pressure further, $T^{\ast}$ is suppressed to a lower temperature and is no longer visible above $\sim$2~GPa. In the high-pressure regime, $\rho(T)$ shows a metallic-like temperature dependence, and superconductivity appears at low temperatures, which will be discussed later (see Fig.~4). As shown in Figs.~2(b) and 2(c), $T^{\ast}$ is more visible in the temperature dependences of $R_{\rm H}(T)$ and $S(T)$ even at $\sim$2~GPa, where $T^{\ast}$ is hidden in the resistivity data. It should be noted that the overall magnitude of $|R_{\rm H}(T)|$ decreases with pressure, which is consistent with the aforementioned variation in $\rho(T)$.
	
	The transport properties of LaOPbBiS$_{3}$ with a lower $T^{\ast}$ show a similar pressure response but are more sensitive to pressure than those of La$_{2}$O$_{2}$Bi$_{3}$AgS$_{6}$. As shown in Fig.~3(a), the semiconducting behavior is rapidly suppressed by applying pressure, and $\rho(T)$ becomes metallic at $\sim$2~GPa. Accordingly, the clear anomaly in $R_{\rm H}(T)$ and $S(T)$ at $T^{\ast}$ [see Figs.~3(b) and 3(c), respectively] is not only shifted toward lower temperatures but also decreases in amplitude. It is noteworthy that $\rho(T)$ in the non-metallic regime follows Mott variable range hopping (VRH) in a wide temperature region, especially below $T^{\ast}$ [see the inset of Fig.~3(a)].
	
	Our central finding is the evolution of superconductivity in both compounds. As shown in Fig.~4, the superconducting transition temperature $T_{\rm c}$ emerging above $\sim$0.5~GPa increases with pressure and reaches a maximum of 2.8~K and 4.6~K for La$_{2}$O$_{2}$Bi$_{3}$AgS$_{6}$ and LaOPbBiS$_{3}$, respectively. The large diamagnetic signal exceeds nearly 60$\%$ of the superconducting shielding fraction and occurs at the temperature at which $\rho(T)$ vanishes, providing strong evidence for the occurrence of superconductivity. Note that the values of $T_{\rm c}$ and $\rho(T)$ at 0.5~GPa in La$_{2}$O$_{2}$Bi$_{3}$AgS$_{6}$ are similar to those of the ambient-pressure superconducting sample \cite{Jha}. Therefore, the appearance of superconductivity at ambient pressure in the Ag-excess sample with a smaller lattice constant is due to the chemical pressure effect.
	 
\begin{figure}[t]
\begin{center}
\includegraphics[width=0.45\textwidth]{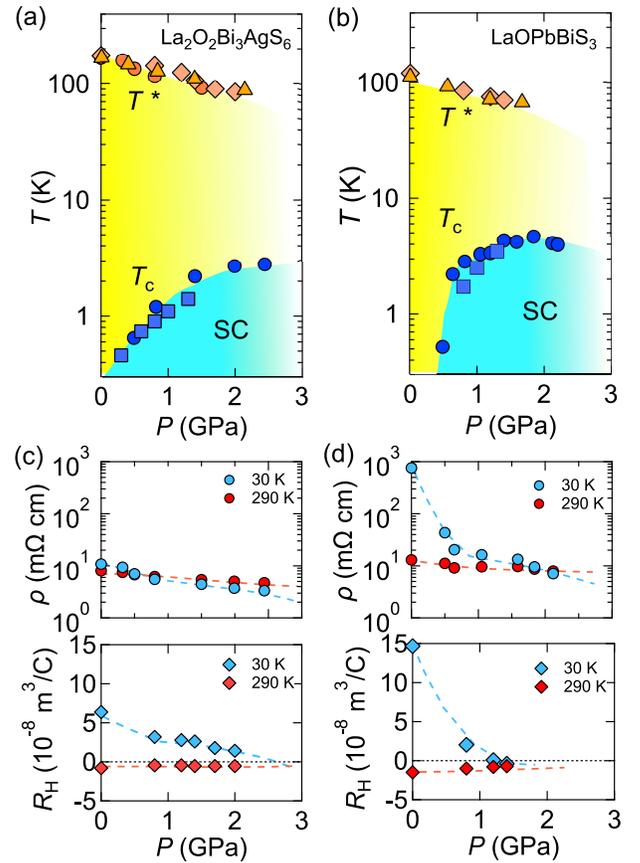}
\end{center}
\caption{Temperature--pressure phase diagram of (a) La$_{2}$O$_{2}$Bi$_{3}$AgS$_{6}$ and (b) LaOPbBiS$_{3}$. Here, the circle, diamond, triangle, and square points are determined from $\rho(T)$, $R_{\rm H}(T)$, $S(T)$, and $\chi_{\rm ac}(T)$, respectively. Pressure dependence of $\rho$ and $R_{\rm H}$ at the temperatures of 30 and 290~K for (c) La$_{2}$O$_{2}$Bi$_{3}$AgS$_{6}$ and (d) LaOPbBiS$_{3}$. The lines are guides for the eye.}
\label{Fig5}
\end{figure}

	The obtained results are summarized in the pressure-temperature phase diagrams of La$_{2}$O$_{2}$Bi$_{3}$AgS$_{6}$ and LaOPbBiS$_{3}$ shown in Fig.~5. While the anomaly at $T^{\ast}$ determined by the transport measurements decreases monotonically with increasing pressure, the signature of superconductivity appears above $\sim$0.5~GPa. Upon increasing the pressure further, $T_{\rm c}$ tends to saturate and exhibits a peak in LaOPbBiS$_{3}$ with a maximum value higher than that obtained in the case of chemical substitution\cite{Jha_2020}. At around the optimum pressure for both compounds, the temperature dependence of $\rho$ and $R_{\rm H}$ becomes weak [see Figs.~5(c) and 5(d)], and the carrier density $n$, determined from $R_{\rm H}$ as $n$ = 1/$|R_{\rm H}e|$, is estimated to be on the order of 1 $\times$ 10$^{21}$ cm$^{-3}$. In this work, we used hydrostatic pressure as a tuning parameter and revealed that La$_{2}$O$_{2}$Bi$_{3}$AgS$_{6}$ and LaOPbBiS$_{3}$ share a few common features: First, the occurrence of pressure-induced superconductivity accompanies the suppression of the semiconductorlike behavior. Second, $T^{\ast}$ lies on both sides of the crossover region from the semiconducting to the metallic state. Third, an opposite trend is found such that $T_{\rm c}$ increases as $T^{\ast}$ decreases upon applying pressure, implying a close relationship between $T_{\rm c}$ and $T^{\ast}$. In a previous report\cite{Jha}, a partial gaplike anomaly $T^{\ast}$, which is suggestive of the CDW transition, was observed in a metallic sample. However, in the phase diagrams obtained in the present work, $T^{\ast}$ extends across the low-pressure semiconductor regime; therefore, we rule out the possibility of a CDW transition driven by the conventional Fermi-surface nesting in the metallic state.
	
	It was proposed that the semiconductorlike behavior in LaOPbBiS$_{3}$ is induced by the local disorder, and this is suppressed by Se substituting the in-plane S site, which serves as the in-plane chemical pressure, similar to other BiS$_{2}$ systems\cite{Mizuguchi_2015}. When hydrostatic pressure is applied, a shrinkage in the in-plane lattice parameters is naturally expected upon compression as a consequence of the weaker disorder. This is consistent with the observation of VRH conduction in LaOPbBiS$_{3}$ and its suppression by  pressure. In this context, the superconducting transition temperature $T_{\rm c}$ in La$_{2}$O$_{2}$Bi$_{3}$AgS$_{6}$ and LaOPbBiS$_{3}$ is enhanced around the crossover region from the strong to weak disorder regime. At present, the origin of the $T^{\ast}$ emerging around the crossover between the semiconductor and metallic states is unclear. Further studies, including detailed structural analyses and theoretical studies, are required to clarify the nature of the anomaly at $T^{\ast}$.
	
\section{conclution}
	In summary, we investigated the effect of pressure on polycrystalline LaOPbBiS$_{3}$ and La$_{2}$O$_{2}$Bi$_{3}$AgS$_{6}$ and established the generic phase diagram of the four-layer-type conducting layer. The enhancement in the pressure-induced superconducting transition temperature $T_{\rm c}$ lies on the way to the suppression of the disorder, which is inferred from the pressure dependence of the transport properties. These results provide insight into the interplay between superconductivity and disorder, which is important to gain a deeper understanding of the underlying physics of BiS$_{2}$-based superconductors.

\section*{ACKNOWLEDGMENTS}
	We thank H.~Kusunose for helpful discussions as well as M.~Hedo, K.~Kitagawa, and Y.~Uwatoko for experimental support and discussions. This work was partially supported by JSPS KAKENHI Grants No. 18H01172, No. 18H04312 (J-Physics), No. 19H01836, and No. 21K03442. The use of the facilities at the Coordinated Center for UEC Research Facilities and the Cryogenic Center at the University of Electro-Communications is appreciated.

\bibliography{Reference}

\end{document}